\documentclass[10pt,twocolumn,english,aps,showpacs,superscriptaddress,groupaddress,floatfix,graphics,graphicx]{revtex4-2}
\usepackage{amsmath,amsfonts,amssymb,epsfig,dcolumn,bm,dsfont,graphics,latexsym,color,graphicx,multirow,ulem}
\usepackage[bookmarks,colorlinks,linkcolor=blue,citecolor=blue,urlcolor=blue]{hyperref}

\let\oldAA\AA

\renewcommand{\AA}{\text{\normalfont\oldAA}}
\newcommand{\ket}[1]{| {#1} \rangle} 
\newcommand{\aver}[1]{\langle {#1} \rangle} 
\begin{document}
\preprint{AIP/123-QED}
\title{Gate-tunable synthetic antiferromagnetism with nonrelativistic spin splitting in a graphene/MnS/graphene heterostructure}
\author{Marko Milivojevi\'c}
\email{marko.milivojevic@savba.sk}
\affiliation{Institute of Informatics, Slovak Academy of Sciences, 84507 Bratislava, Slovakia}
\affiliation {Faculty of Physics, University of Belgrade, 11001 Belgrade, Serbia}
\author{Martin Gmitra}
\affiliation{Institute of Physics, Pavol Jozef \v{S}af\'{a}rik University in Ko\v{s}ice, 04001 Ko\v{s}ice, Slovakia}
\affiliation{Institute of Experimental Physics, Slovak Academy of Sciences, 04001 Ko\v{s}ice, Slovakia}
\affiliation{New Technologies Research Centre, University of West Bohemia, Univerzitní 8, CZ-301 00 Pilsen, Czech Republic}
\begin{abstract}
We propose encapsulating type-A antiferromagnetic semiconductors between graphene layers to realize a gate-tunable synthetic antiferromagnet with nonrelativistic spin splitting, enabling efficient spintronic transport via graphene.
Density functional theory calculations and tight-binding models of graphene/MnS/graphene heterostructure reveal that gate-tuning of the heterostructure breaks top/bottom graphene equivalence, inducing opposite ferromagnetic proximity exchange that lifts spin degeneracy to yield nonrelativistic spin splitting at the Fermi level, dominating over relativistic effects.
The induced effects manifest as conductance dips in spin-resolved transport through proximitized graphene nanoribbons, observable as giant magnetoresistance within a narrow energy window around the Fermi level.
Our graphene/type-A antiferromagnetic heterostructure, a readily synthesizable platform incorporating antiferromagnets with nonrelativistic spin splitting, paves the way for gate-manipulated, low-dimensional antiferromagnetic devices.
\end{abstract}
\maketitle
\section{Introduction}
Altermagnets and a more general class of antiferromagnets with nonrelativistic spin splitting (NRSS  AFMs)~\cite{HYK19,HYK20,YWL+20,YWL+21,SSJ22I,SSJ22} represent a widely discussed class of magnetic order, distinct from ferromagnets and conventional AFMs. They feature collinear, alternating spin sublattices that yield zero net magnetization (like AFMs), while having crystal rotations or mirrors connecting opposite-spin motif pairs~\cite{YZ23}. This produces wavevector-dependent spin splitting in altermagnets, or even at the Brillouin 
zone center in a recently discovered class of NRSS AFMs~\cite{YGR24}. These features enable diverse spintronic phenomena like anomalous~\cite{ALK24,SD25} and skyrmion~\cite{JZC+24} Hall effect,  spin filtering~\cite{SST25}, spin-polarized currents~\cite{HHH+25,SYC25},
Majorana zero modes~\cite{HGC25}, and ultrafast dynamics~\cite{EAF+25,ZH25,WLH+25}, ideal for spintronics~\cite{ZFS04,FME+07}.

For spintronics applications, two-dimensional (2D) materials outperform their three-dimensional (3D) counterparts due to stronger spin-orbit coupling (SOC)~\cite{ZCS11}, atomically thin~\cite{NGM+04,MLH+10} structures that enable precise electrostatic control, and  integration into van der Waals (vdW) heterostructures~\cite{GG13,Sierra2021}. 
In 2D systems, altermagnets face stricter symmetry constraints than in 3D~\cite{ZZ24}, resulting in only a limited number of candidate materials~\cite{MOP+24,SO24,LLS+25}. None of these, however, have been experimentally realized, prompting alternative strategies to realize 2D altermagnets. Recent approaches include (i)~functionalizing conventional AFMs with external stimuli such as electric fields~\cite{MHS23} or strain~\cite{ZDZ25}; and (ii) engineering vdW heterostructures with tailored altermagnetic properties \cite{ZGL+25,MHS+25,SZY+25}.
Going beyond altermagnets, a novel class of NRSS AFMs lacks crystal symmetries connecting opposite-spin sublattices~\cite{YGR24}, allowing to
broaden the scope of eligible 2D AFM materials for innovative devices.

\begin{figure}[t]
    \centering
 \includegraphics[width=0.99\linewidth]{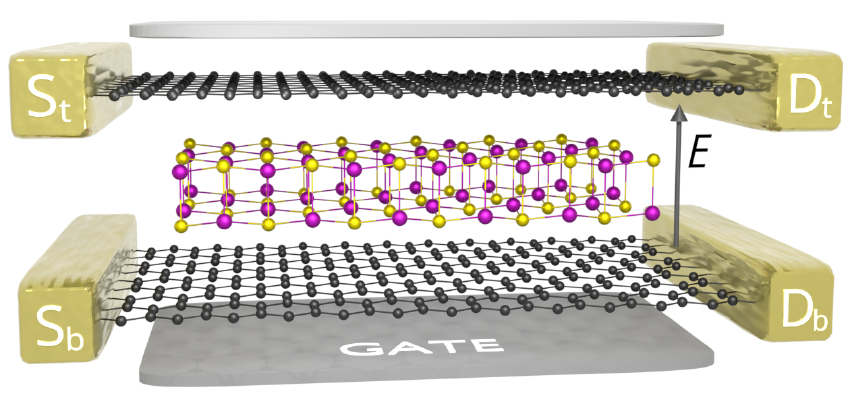}
    \caption{Schematic view of the transport device that
implements the synthetic vdW NRSS AFM formed by
encapsulating a monolayer of MnS (type-A AFM
semiconductor) between graphene layers.
The top and bottom graphene layers are independently contacted by separate source-drain pairs,
$\mathrm{S_t}$-$\mathrm{D_t}$ and
$\mathrm{S_b}$-$\mathrm{D_b}$, respectively.
The central region consists of graphene/MnS/graphene under an applied perpendicular electric field $E$ modulated by a gate voltage.}
    \label{fig:transport}
\end{figure}

By exploiting this novel class of NRSS AFMs~\cite{YGR24},
we demonstrate electrically switchable NRSS in a gated
graphene/MnS/graphene heterostructure (FIG.~\ref{fig:transport}).
An out-of-plane electric field induces NRSS in monolayer MnS, a type-A AFM semiconductor, by breaking the inversion symmetry between spin sublattices.
Since monolayer MnS is a semiconductor with a sizable band gap~\cite{SIC22}, it cannot serve as an efficient spin-transport medium on its own.
By encapsulating MnS between graphene layers, we transfer the proximity-induced NRSS into graphene, obtaining an active spin-transport channel with gate-tunable carrier density, thereby realizing a gate-switchable synthetic vdW NRSS AFM.
Focusing on the Fermi level, where only graphene bands contribute, our tight-binding model of graphene with density functional theory-extracted parameters reveals gate-tunable doping and sublattice-resolved proximity exchange. We uncover clear proximity signatures of NRSS AFM in the ballistic transport regime of graphene nanoribbons, manifesting as giant magnetoresistance.
These findings demonstrate strong tunability of the device within an electrically controlled doping window.

\section{Structural model and electronic structure}
Layered MnS AFM contains two atomic layers forming a trigonal crystal structure, with 
P$\overline{3}$m1 space group (No. 164)~\cite{LH25}. 
Magnetic order is type-A AFM, where the ferromagnetically ordered Mn atoms within each atomic layer are AFM coupled.
Owing to the presence of inversion symmetry between spin sublattices~\cite{SSJ22}, bands of MnS are spin degenerate~\cite{LH25}. However, NRSS in MnS can be induced and tuned by a perpendicular electric field, since it breaks the inversion symmetry between spin sublattices~\cite{LH25}.

The studied graphene/MnS/graphene heterostructure is obtained by intercalating an AA-stacked bilayer graphene by a monolayer of MnS.
Since both MnS and AA-stacked bilayer graphene individually possess
inversion symmetry, their symmetric encapsulation preserves the
inversion symmetry between the spin sublattices of the full
heterostructure, thus forbidding the removal of spin degeneracy in the absence of an external electric field.
The lattice parameter of graphene is taken as $a_0=2.46~{\rm\AA}$, while for MnS, the latice parameters is equal to $a_{\rm MnS}=4.11~{\rm\AA}$~\cite{SIC22}. The supercell is constructed by straining graphene, described by the relative strain $\delta=0.244\%$, using which the strained lattice parameter of graphene $a_0^{\rm str}$ can be described as $a_0^{\rm str}=(1+\delta[\%]/100)a_0$.  Using the strained lattice vectors of graphene ${\bm a}_1=a_0^{\rm str}{\bm e}_x$ and ${\bm a}_2=a_0^{\rm str}(\cos{(2\pi/3)}{\bm e}_x+\sin{(2\pi/3)}{\bm e}_y)$ we can define lattice vectors of the heterostructure $(5,0)$ and $(0,5)$ as $(5,0)=5{\bm a}_1$ and $(0,5)=5{\bm a}_2$. The total number of atoms in the heterostructure is 137, whereas the top and bottom graphene consist of 50 atoms each.
Details of the ab-initio calculations can be found in the Supplemental Material~\cite{SUPP} (see also
Refs.~\cite{QE1,QE2,PBE,pslib,PAW,MP89,G06,BCF+08,PZ81,B99}
therein).

Nonrelativistic spin degeneracy is broken when the perpendicular electric field is applied, transforming the graphene/MnS/graphene heterostructure into a synthetic vdW NRSS AFM. Crucially, averaged magnetic moments on top/bottom Mn atoms remain nearly identical for electric fields magnitudes up to 1~V/nm, confirming zero net magnetization and excluding ferrimagnetism. We prove this via density functional theory nonrelativistic spin-collinear calculations, see Section~2.1 of Supplementary Materials (SM) and Fig.~S1~\cite{SUPP}, which demonstrate field-tunable spin splitting in the top/bottom graphene layers arising from gate-induced NRSS in MnS, thereby ruling out relativistic SOC origins.

Having established the nonrelativistic origin of the proximity-induced spin splitting in graphene bands, we now focus on relativistic density functional theory (DFT) calculations of the studied heterostructure to explore the role of SOC in the system, which can have a sizable effect. To this end, we perform a series of DFT calculations with different electric field strengths applied perpendicular to the heterostructure plane.
\begin{figure}[t]
    \centering
    \includegraphics[width=0.9\linewidth]{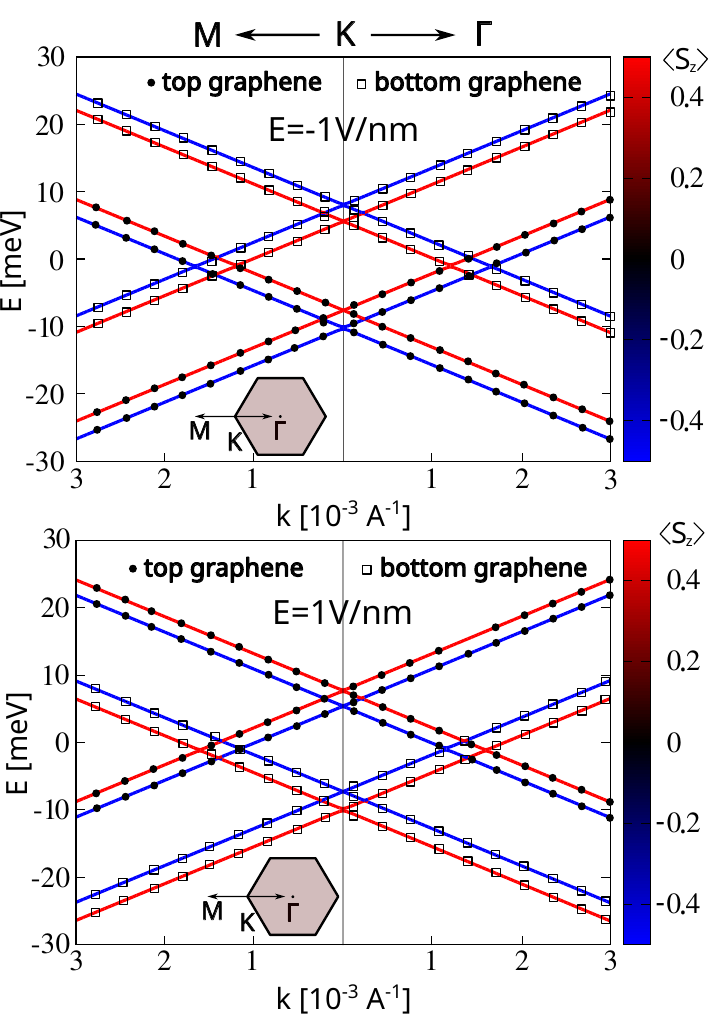}
    \caption{Calculated band structure of graphene/MnS/graphene heterostructure under an applied out-of-plane electric field of $E=\pm 1$~V/nm (relativistic case), plotted along the high symmetry points MK$\Gamma$ in the graphene Brillouin zone. The plot shows a narrow energy window of $[-30,\,30]$~meV around the Fermi level, in which the bands are confined to the (proximitized) graphene layers. The color scale indicates the expectation value of the spin 
    $z$-component, revealing nearly perfect 
    $\pm 1/2$ spin polarization along $z$. Black circles (squares) denote projections onto top (bottom) graphene orbitals, highlighting the electric-field-tunable shift of the Dirac cones in the top (bottom) layer.}
    \label{fig:band-structure}
\end{figure}
In FIG.~\ref{fig:band-structure} we present the band structure of the studied heterostructure under an applied electric field of $E=\pm1$ V/nm in the narrow energy window $[-30,\,30]$\,meV around the Fermi level, together with the orbital projection of the given bands to orbitals of top graphene (black circles) and bottom graphene (empty squares). The results indicate that the band structure near the Fermi level fully belongs to graphene, which is expected due to the 
semiconducting MnS~\cite{SIC22}. On the other hand, bands belonging to top and bottom graphene are fully decoupled, as expected, due to their spatial separation by the intervening MnS layer.
Additionally, we see that the obtained band structure is highly dependent on the applied electric field strength: for $E=-1(+1)$~V/nm, the top graphene is below (above) the Fermi level, while the bottom layer behaves oppositely, indicating electric-field-induced modulation of the graphene band structure.
Additionally, the color scale represents the expectation value of the spin $z$-component, revealing nearly perfect $\pm$1/2 spin polarization in the $z$-direction.

\begin{table}[t]
\caption{Parameters of the effective tight-binding model of top and bottom graphene (relativistic case) within graphene/MnS/graphene heterostructure for electric field strengths $E=\pm1$\,V/nm.} \label{TAB:RELparametersMAIN}
\centering
\small
\setlength{\tabcolsep}{7pt}
\renewcommand{\arraystretch}{1.0}
\begin{tabular}{rrr}
\hline\hline
$E=-1$~V/nm & top graphene& bottom graphene \\ \hline
$v_{\rm F}$ [$10^{6}${\rm m/s}]        &0.820  &0.819\\
$\mu$\;[{\rm meV}]                     &-8.911 &6.820\\
$\Delta$\;[{\rm meV}]                  &-0.001 &-0.017\\
$\lambda_{\rm I}^{\rm A}$\;[{\rm meV}] &0.116  &-0.014 \\
$\lambda_{\rm I}^{\rm B}$\;[{\rm meV}] &0.073  &0.042\\
$\lambda_{\rm R}$\;[{\rm meV}]         &0.003  &0.002 \\
$\Delta_{\rm A}$\;[{\rm meV}]          &1.312&-1.118 \\
$\Delta_{\rm B}$\;[{\rm meV}]          &1.278&-1.279\\\hline\hline
$E=1$~V/nm &  top graphene & bottom graphene \\ \hline
$v_{\rm F}$ [$10^{6}${\rm m/s}]        &0.820&0.818\\
$\mu$\;[{\rm meV}]                     &6.564&-8.597\\
$\Delta$\;[{\rm meV}]                  &-0.168&0.066\\
$\lambda_{\rm I}^{\rm A}$\;[{\rm meV}] &-0.004&-0.054\\
$\lambda_{\rm I}^{\rm B}$\;[{\rm meV}] &-0.025&-0.032\\
$\lambda_{\rm R}$\;[{\rm meV}]         &-0.003&0.003\\
$\Delta_{\rm A}$\;[{\rm meV}]          &1.123&-1.277\\
$\Delta_{\rm B}$\;[{\rm meV}]          &1.126&-1.382\\\hline\hline
\end{tabular}
\end{table} 
\section{Tight-binding modeling}\label{sectionIII}
 To clarify the proximity-induced spin physics in graphene, we employ a tight-binding model of graphene bands around the Fermi level, which we describe by two dependent (top and bottom) graphene monolayers, 
$H_{\rm gr/MnS/gr}^{\rm eff}=H_{\rm gr}^{\rm top}+H_{\rm gr}^{\rm bottom}$, where top and bottom graphene layer have the same structural form, albeit different effective parameters to be determined by fitting the model Hamiltonian of top/bottom graphene to DFT data. For graphene influenced by both the proximity-exchange coupling and proximity-induced spin-orbit coupling, the effective Hamiltonian can be written as
$H_{\rm gr}=H_0+H_{\rm I}+H_{\rm{R}}+H_{\rm{ex}}$, 
where we exploit the geometric (nonmagnetic) ${\bf C}_{3{\rm v}}$
symmetry of the heterostructure to model the SOC~\cite{Kochan2017}. Here $H_0$ represents the orbital Hamiltonian, described in terms of the sublattice-dependent on-site 
potential, $\sum_{i=\rm{A},\rm{B}}\sum_{\sigma}(\mu+\Delta (-1)^{\delta_{\rm{B},i}})c_{i\sigma}^{\dagger}c_{i\sigma}$,
equal to $\mu\pm\Delta$ on sublattice A/B of graphene, where $\mu$ is the chemical potential, $\Delta$ is the stagerred potential, and $c_{i\sigma}^{\dagger}$ ($c_{i\sigma}$) is the creation (anihilation) operator on site $i=\rm{A},\rm{B}$ with spin $\sigma=\uparrow\downarrow$. Furthermore, the orbital Hamiltonian consists of nearest-neighbor Hamiltonian with hopping $t$, $-t \sum_{\aver{m,n}}\sum_{\sigma}c_{n\sigma}^{\dagger}c_{m\sigma}$, which can be connected to the Fermi velocity $v_{\rm F}$ through the relation $v_{\rm F}=a t\sqrt{3}/2\hbar$, where $a$ is the lattice constant of graphene (see SM~\cite{SUPP}). 
The second term, $H_{\rm I}=\sum_{\alpha={\rm A,B}}\sum_{\aver{\aver{m,n}}}\frac{{\rm i}\lambda_{\rm I}^{\alpha}}{3\sqrt{3}}\sum_{\sigma}\nu_{m,n}[s_z]_{\sigma\sigma}c_{m\sigma}^{\dagger}c_{n \sigma}$, describes the intrinsic SOC, where the sum goes over the next-nearest-neighbors interaction described in terms of the sublattice-dependent $\lambda_{\rm I}^{\rm{A}/\rm{B}}$ parameters and the sign factor $\nu_{m,n}$ that has the value 1 ($-1$) when the next-nearest-neighbor hopping from site $m$ to site $n$ via the common nearest-neighbor encloses a clockwise (counterclockwise) path.  Next, $H_{\rm R}$, describes the nearest-neighbor Rashba SOC term, $H_{\rm{R}}=\frac{2{\rm i}\lambda_{\rm R}}{3}\sum_{\aver{m,n}}\sum_{\sigma\neq\sigma'}
[{\bf s}\times {\bf d}_{m,n}]_{\sigma\sigma'}^z c_{m\sigma}^{\dagger}c_{n \sigma'}$, where ${\bf s}$ is the vector of Pauli matrices, $\lambda_{\rm R}$ represents the Rashba SOC strength, and ${\bf d}_{m,n}$ is the unit vector in the horizontal plane pointing from lattice site $n$ to the nearest-neighbor site $m$. 
Finally, the presence of proximity-induced magnetization
in graphene induced by the antiferromagnetic
monolayer~\cite{AHK+21} is modeled by the
sublattice-resolved exchange Hamiltonian, $H_{\rm ex}=\sum_{i=\rm{A},\rm{B}}\sum_{\sigma}\Delta_i [s_z]_{\sigma\sigma} c_{i\sigma}^{\dagger}c_{i\sigma}$, where $\Delta_{\rm A}$ and $\Delta_{\rm B}$ of the top and bottom graphene monolayers are parameters to be determined. 

For the previously analyzed examples of the graphene/MnS/graphene heterostructure subjected to the electric field strengths $E=\pm1$~V/nm, in Table~~\ref{TAB:RELparametersMAIN} we provide parameters of the effective tight-binding model of top and bottom graphene, obtained after fitting the DFT data to the model in the vicinity of K and K' points of graphene, 
see Section S2 of SM for more details~\cite{SUPP}. The presented parameters reveal the ferromagnetic proximity exchange in both top and bottom graphene layers ($\Delta_{\rm A/B}^{\rm t}>0$, $\Delta_{\rm A/B}^{\rm b}<0)$, however with opposite sign due to the proximity effect of the nearest ferromagnetic MnS layer. When compared to the proximity-induced SOC parameters, exchange interaction is much stronger than both the Rashba (which can be safely neglected) and the intrinsic SOC (representing a small correction to the exchange term), suggesting the dominant magnetic proximity effect in graphene monolayers when interfaced with magnetic materials~\cite{ZGF+16,GPI+25}. 
These results indicate ferromagnetic proximity~\cite{WTS+15,XSK+18,YGH+25} effects in top/bottom graphene layers, whereas the opposite signs of the exchange parameters give rise to overall AFM behavior of the heterostructure; combined with different chemical potentials in the two layers, this yields NRSS and justifies the term synthetic NRSS AFM for the studied heterostructure. In addition, the chemical potentials exhibit sign changes upon field reversal ($\mu^{\rm t/b}>0$ for $E=\pm1$~V/nm and vice versa), consistent with the electron/hole doping of top/bottom graphene seen in FIG.~\ref{fig:band-structure}. These ferromagnetic proximity effects—accompanied by electric-field-controlled doping of the graphene monolayers and weak SOC—hold generally across field strengths, as confirmed by effective parameters fitted for $-1<E<1$~V/nm in steps of 0.25~V/nm (see Table~S3 and Section S2.2 of SM~\cite{SUPP}).\\
\section{Transport signatures}
To probe the transport signatures of proximity-induced NRSS in graphene, we study zigzag nanoribbons where pristine source (S) and drain (D) regions are interfaced with a central (C) synthetic NRSS AFM (FIG.~\ref{fig:transport}).  Using the Green's function method, the conductivity of proximitized graphene nanoribbon was calculated using the Landauer-B{\" u}ttiker formula~\cite{D97}. 
The tight-binding parameters for the
nanoribbon Hamiltonian were extracted from the electronic structure of the full 2D periodic graphene/MnS/graphene heterostructure. 
The nanoribbon geometry, therefore, serves as a two-terminal transport model device.
In a realistic device, separate source-drain contact pairs, S$_{\rm t}$-D$_{\rm t}$ and S$_{\rm b}$-D$_{\rm b}$, give direct experimental access to the layer and spin-resolved conductance as discussed below.

The zero-temperature spin-resolved conductance $G_{\sigma}(\varepsilon)$ at energy $\varepsilon$, $\sigma=\uparrow\downarrow$, of the graphene nanoribbon is given by
$G_{\sigma}(\varepsilon)=\frac{e^2}{\hbar}{\rm Tr}[\Gamma_{\rm S}(\varepsilon)G^{\rm r}(\varepsilon)\Gamma_{\rm D}(\varepsilon)
G^{\rm a}(\varepsilon)]_{\sigma}$, where 
$G^{\rm r(a)}(\varepsilon)=[\varepsilon\,{\rm I}-H_{\rm C}-\Sigma_{\rm S}^{\rm r(a)}(\varepsilon)-\Sigma_{\rm D}^{\rm r(a)}(\varepsilon)]^{-1}$
is the retarded (advanced) Green's function matrix of the central part, $H_{\rm C}$ is the Hamiltonian of the central region, while $\Sigma_{\rm S(D)}^{\rm r(a)}(\varepsilon)$ is the retarded (advanced) self-energy due to the source (drain), and broadening matrices $\Gamma_{\rm S(D)}(\varepsilon)$ are defined as
$\Gamma_{\rm S(D)}(\varepsilon)={\rm i}[\Sigma_{\rm S(D)}^{\rm r}(\varepsilon)-\Sigma_{\rm S(D)}^{\rm a}(\varepsilon)]$.
Within this approach, the difficulty of treating an infinite system is shifted to the calculation of the self-energies, $\Sigma_{\rm S}=\tau_{\rm CS}g_{\rm S}\tau_{\rm SC}$, $\Sigma_{\rm D}=\tau_{\rm CD}g_{\rm D}\tau_{\rm DC}$, where $\tau_{\rm CS/SC/DC/CD}$ are the Hamiltonian coupling matrices between the central region and source and drain, while the surface Green’s functions of the left electrode $g_{\rm S}$ and the right electrode $g_{\rm D}$ can be calculated iteratively with the renormalization-decimation algorithm as described in~\cite{LLR84,LLR85,TZS+19}.

One possibility of quantifying the exchange proximity effect induced in top/bottom graphene is by calculating spin-resolved magnetoresistance, defined as a change in top/bottom graphene's conductance due to the proximity-induced exchange field, 
\begin{equation}
{\rm MR}^{\rm t/b}_\sigma(\varepsilon,E)=\left(
    \frac{G^{\rm t/b}_{\sigma}(\varepsilon,0)}
    {G^{\rm t/b}_{\sigma}(\varepsilon,E)}-1\right) 100[\%],
\end{equation} 
where $G^{\rm t/b}_{\sigma}(\varepsilon,0)$ represents spin-resolved conductance of ideal top/bottom graphene, whereas $G^{\rm t/b}_{\sigma}(\varepsilon,E\neq0)$
corresponds to the conductance of proximitized graphene. In the case of ideal graphene (without the proximity effect), conductance in the wide energy range around the Fermi level~\cite{YZS+09} is equal to 1 (in the units of $e^2/\hbar$), $G^{\rm t/b}_{\sigma}(\varepsilon,0)=1$. Thus, a rise in magnetoresistance arises from decreases in proximity-induced conductance at certain energies. 
Our numerical analysis, using the tight-binding model with parameters given in Table~\ref{TAB:RELparametersMAIN}, indicates that sizable changes in conductance appear in the short energy window around the Dirac cone of top/bottom graphene. Thus, we will discuss our results in the energy window $[-11,11]$~meV around the Fermi level (of S and D).
\begin{figure}[t]
    \centering
    \includegraphics[width=0.99\linewidth]{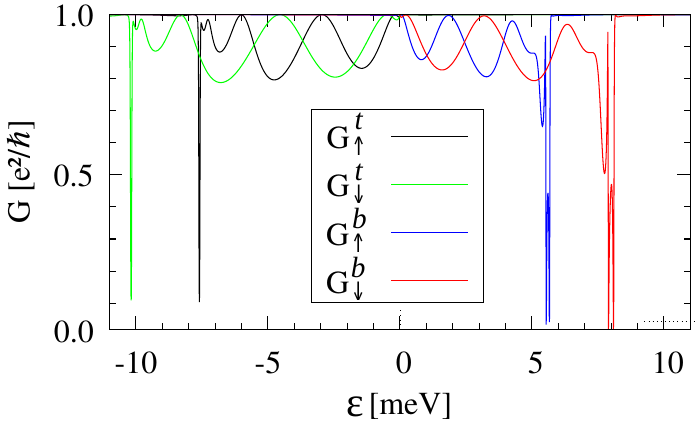}
    \caption{Spin and layer resolved conductance $G^{\rm t/b}_{\sigma}$
($\sigma = \uparrow, \downarrow$) as a function of energy
$\varepsilon$ for a zigzag graphene nanoribbon of length
48~\AA\ and width 17.8~\AA, with the central region
proximitized by MnS under an applied electric field
$E = -1$~V/nm. The spin-up (black) and spin-down (green) conductance channels
of the top graphene layer, $G^{\rm t}_{\uparrow}$ and
$G^{\rm t}_{\downarrow}$, exhibit sharply defined local minima at
negative energies, reflecting the hole doping $\mu^{\rm t} < 0$
of the top layer. The spin-up (blue) and spin-down (red) conductance channels
of the bottom graphene layer, $G^{\rm b}_{\uparrow}$ and
$G^{\rm b}_{\downarrow}$, display analogous minima at positive
energies, consistent with the electron doping $\mu^{\rm b} > 0$
of the bottom layer. Within each layer, the spin-up and spin-down dips are shifted
relative to each other due to the sublattice-resolved exchange parameters $\Delta^{\rm t/b}_{\rm A/B}$, providing a direct transport signature of the proximity-induced nonrelativistic spin splitting.
}
    \label{fig:MR}
\end{figure}

In FIG.~\ref{fig:MR}, we present the spin- and layer-resolved conductance dependency on energy $\varepsilon$ of the graphene nanoribbon with zigzag edges having a length of 48\,$\AA$ and a width of 17.8\,$\AA$, assuming an applied electric field strength of $E=-1$~V/nm. 
One sees two local minima in conductance at positive/negative energies, corresponding to signals from the bottom/top graphene, respectively. 
To understand the influence of different parameters that affect conductance, we first notice that intrinsic SOC parameters for all studied cases (effective parameters are given in Tab.~S2 of supplementary materials), $\lambda_{\rm I}^{\rm A/B}$ have a negligible effect on $G_{\sigma}^{\rm t/b}$, except affecting depth of $G_{\sigma}^{\rm t/b}$ conductance dips, appearing roughly at  $\mu^{\rm t/b}\pm \Delta_{\rm A}^{\rm t/b}$
(since roughly $\Delta_{\rm A}^{\rm t/b}$ is equal to $\Delta_{\rm B}^{\rm t/b})$. 
We also note that the diagonal elements of the Hamiltonian $H_{0}+H_{\rm ex}$ in the basis $\{\ket{A^{\rm t/b}\uparrow},\ket{A^{\rm t/b}\downarrow},
\ket{B^{\rm t/b}\uparrow},\ket{B^{\rm t/b}\downarrow}\}$, are equal to 
$\mu^{\rm t/b}+\Delta^{\rm t/b}+\Delta_{\rm A}^{\rm t/b},
\mu^{\rm t/b}+\Delta^{\rm t/b}-\Delta_{\rm A}^{\rm t/b},
\mu^{\rm t/b}-\Delta^{\rm t/b}+\Delta_{\rm B}^{\rm t/b},
\mu^{\rm t/b}-\Delta^{\rm t/b}-\Delta_{\rm B}^{\rm t/b}$,
respectively. Due to negligible Rashba SOC, the spin-down and spin-up channels are not coupled. In addition to this, the dominant energy scale is given by $\mu^{\rm t/b}$, following the exchange parameters, $\Delta_{\rm A}^{\rm t/b}$ and $\Delta_{\rm B}^{\rm t/b}$. In contrast, the staggered potential $\Delta^{\rm t/b}$ is on the same energy scale as the energy difference $\Delta^{\rm t/b}\propto|\Delta_{\rm A}^{\rm t/b}-\Delta_{\rm B}^{\rm t/b}|$. Taking all this into account, one can conclude that each conductance channel $G_{\sigma}^{\rm t/b}$, $\sigma=\uparrow\downarrow$, can have two dips at $\mu^{\rm t/b}+\Delta^{\rm t/b}+\sigma\Delta_{\rm A}^{\rm t/b}$ and
$\mu^{\rm t/b}-\Delta^{\rm t/b}+\sigma\Delta_{\rm B}^{\rm t/b}$.
Thus, depending on the difference between $\Delta^{\rm t/b}+\sigma\Delta_{\rm A}^{\rm t/b}$
and $-\Delta^{\rm t/b}+\sigma\Delta_{\rm B}^{\rm t/b}$, one can see zero broadening of dips, as in the case given in FIG.~\ref{fig:MR} for top graphene, whereas for bottom graphene, finite size broadening (or two clearly distinguishable dips) of a single $G_{\sigma}^{\rm b}$ channel, suggests that either $\Delta^{\rm b}$ is nonzero or difference between $\Delta_{\rm A}^{\rm b}$
and $\Delta_{\rm B}^{\rm b}$ is not negligible (or in some cases both). 
This simple line of reasoning is backed 
by the effective parameters given in Table~\ref{TAB:RELparametersMAIN} for the studied electric field strength: for the case of top graphene, $\Delta^{\rm t}$ is negligible, as well as the difference between $\Delta_{\rm A}^{\rm t}$ and
$\Delta_{\rm B}^{\rm t}$. On the other hand, a noticeable difference of 0.161 meV between $\Delta_{\rm A}^{\rm b}$ and
$\Delta_{\rm B}^{\rm b}$ is responsible for finite size broadening of $G_{\sigma}^{\rm b}$ dips, whereas asymmetry between the $G_{\uparrow}^{\rm b}$ and 
$G_{\downarrow}^{\rm b}$ suggest nonzero staggered potential.

Consequently, the maximal magnetoresistivity arises precisely at these conductance dips of the proximitized top/bottom graphene layers, where pronounced drops in $G^{\rm t/b}_{\sigma}(\varepsilon,E)$ relative to the ideal value of 1 ($e^2/\hbar$) yield the largest ${\rm MR}^{\rm t/b}_{\sigma}$ enhancements.
At the conductance dip energies, $G^{\rm t/b}_\sigma(\varepsilon,E)$ drops well below the ideal graphene value of 1  ($e^2/\hbar$), so that the magnetoresistance ratio MR$^{\rm t/b}_\sigma(\varepsilon,E)$ becomes very large, far exceeding the proximity magnetoresistance values of up to 100\% reported in analogous graphene-based systems~\cite{SHW+19}.
Positioned approximately at $\mu^{\rm t/b} \pm \Delta^{\rm t/b} \pm \sigma \Delta_{\rm A}^{\rm t/b}$ and modulated by differences between $\Delta_{\rm A}^{\rm t/b}$ and $\Delta_{\rm B}^{\rm t/b}$ or nonzero staggered potentials, these features can exhibit giant values, depending on the electric field strength, nanoribbon length, and sublattice exchange parameters. Finite-size broadening or near-degeneracies amplify these pronounced minima, driving extreme MR signals within the narrow energy window around the Fermi level. In SM~\cite{SUPP}, we present conductance results for different nanoribbon lengths and electric field strengths, further supporting our conclusions and underscoring the potential for gate-tunable spintronic applications in graphene-based antiferromagnetic heterostructures.

The predicted conductance dips rely on the ballistic
transport regime, and it is therefore important to discuss the robustness of these features against disorder.
In general, even a single lattice defect in a graphene nanoribbon can produce conductance dips on the energy scale of the hopping parameter $t \sim \mathrm{eV}$~\cite{YZS+09}, far exceeding the meV-scale proximity splittings discussed here.
Nevertheless, meV-scale proximity-induced spin splittings have been
clearly resolved in graphene heterostructures~\cite{BST+20,HSI+20,YGH+25,GPI+25}, 
demonstrating that meV-scale proximity effects are within experimental transport reach. 
Also, graphene devices of this quality are furthermore known to support micrometer-scale ballistic transport at room temperature~\cite{MGM+11}, consistent with the ballistic regime assumed in our model.
Crucially, even in the presence of residual disorder, the proximity-induced dips carry a distinct experimental signature: their positions, set by $\mu^{\rm t/b}$ and $\Delta^{\rm t/b}_{\rm A/B}$, shift systematically with the applied electric field, interchanging between the top and bottom graphene layers upon the field reversal $E \to -E$, whereas disorder-induced dips are pinned in energy and do not respond to the gate.
The differential conductance measured as a function of gate voltage makes it possible to distinguish between the proximity signatures and the disorder background.

We note that the sharp conductance dips in Fig.~\ref{fig:MR} are predicted in the zero-temperature limit. 
At finite temperatures, thermal broadening can smear these features, reducing the magnetoresistance signal.
A quantitative assessment of finite-temperature effects is left for future investigation.
Nevertheless, we emphasize that the key experimental signature of the predicted effect, the gate-controlled shift of the conductance dip pattern when changing the electric field strength and the electron/hole doping of top/bottom graphene upon electric field reversal, is a qualitative feature connected to the sign change of $\mu^{\rm t/b}$ and $\Delta^{\rm t/b}_{\rm A/B}$, and remains distinguishable at finite temperatures, as long as the exchange splitting and chemical potential shift are resolvable within the thermally broadened window around the Fermi level.

Furthermore, zigzag nanoribbons have been extensively studied in the context of edge magnetism \cite{Yazyev2010:RPP}. 
However, in realistic systems, magnetic edge states may be absent due to structural or chemical modifications \cite{Kurebayashi2022}. 
Experimental observations \cite{Joly2010:PRB} and the functionalization of spin-polarized edge states \cite{Blackwell2021:Nature} provide only indirect evidence for their existence. 
By contrast, the proximity-induced exchange fields $\Delta^{\rm t/b}_{\rm A/B}$ act uniformly on every site of the graphene lattice, including both bulk and edge atoms, and are therefore expected to dominate over the edge magnetization.

\section{Conclusions}
We demonstrated that the symmetric graphene/MnS/graphene heterostructure realizes a synthetic NRSS AFM, where an out-of-plane electric field induces tunable NRSS in the initially spin sublattice inversion-symmetric MnS type-A AFM. Density functional theory-parametrized tight-binding calculations reveal that gate-controlled exchange proximity dominates over SOC, producing sharp conductance dips in spin-resolved nanoribbon transport where only the central region is proximitized. The resulting gate‑ and length‑dependent increases in magnetoresistance occur in a narrow window around the Fermi level, enabling electrically switchable spin-dependent signals in graphene-based synthetic NRSS AFMs.

\acknowledgments
 We thank Srdjan Stavri{\' c} and Jan Kune{\v s} for fruitful discussions. Research results was obtained using the computational resources procured in the national project National competence centre for high performance computing (project code: 311070AKF2) funded by European Regional Development Fund, EU Structural Funds Informatization of society, Operational Program Integrated Infrastructure.
 M.M. acknowledges the financial support by the EU NextGenerationEU through the Recovery and Resilience Plan for Slovakia under Project No. 09I02-03-V01-00012, by the APVV grant APVV-23-0430, and VEGA grants 2/0081/26 and 2/0133/25.
M.G.~acknowledges funding by the EU NextGenerationEU through the Recovery and Resilience Plan for Slovakia under the project No. 09I05-03-V02-00071, the Ministry of Education, Research, Development and Youth of the Slovak Republic, provided under Grant No. VEGA 1/0104/25, the Slovak Academy of Sciences project IMPULZ IM-2021-42, and support of the QM4ST project funded by Programme Johannes Amos Commenius, call Excellent Research (Project No. CZ.02.01.01/00/22\_008/0004572).

{\it Author Contributions.}
M.M. conceived and led the project, performed all calculations,
analyzed the results, and wrote the manuscript.
M.G. contributed through critical reading of the
manuscript and scientific discussions.

\bibliography{biblio}

\end{document}


\date{}
\maketitle

\noindent
\textsuperscript{1}Institute of Informatics, Slovak Academy of Sciences, 84507 Bratislava, Slovakia\\
\textsuperscript{2}Faculty of Physics, University of Belgrade, 11001 Belgrade, Serbia\\
\textsuperscript{3}Institute of Physics, Pavol Jozef \v{S}af\'{a}rik University in Ko\v{s}ice, 04001 Ko\v{s}ice, Slovakia\\
\textsuperscript{4}Institute of Experimental Physics, Slovak Academy of Sciences, 04001 Ko\v{s}ice, Slovakia\\
\textsuperscript{5}New Technologies Research Centre, University of West Bohemia, Univerzitní 8, CZ-301 00 Pilsen, Czech Republic\\
\textsuperscript{*}Email: marko.milivojevic@savba.sk

\section{Details of the DFT calculation}
We perform the electronic structure calculation of the graphene/Mn$_2$S$_2$/graphene heterostructures using Density Functional Theory (DFT) as implemented in the plane wave code Q{\sc{uantum}} ESPRESSO (QE)~\cite{QE1,QE2}. The relaxation of both structures, was performed using the Perdew-Burke-Ernzerhof functional~\cite{PBE} within the projector augmented-wave method~\cite{pslib,PAW}. The kinetic energy cut-offs for the wave function and charge density were chosen to be 55~Ry and 326~Ry, respectively. Additionally, Methfessel–Paxton energy level smearing~\cite{MP89} of 1~mRy was used, and $6\times 6$ $k$-points mesh for the irreducible part of the Brillouin zone sampling were used for self-consistent calculations. 
Furthermore, to accurately describe strongly
correlated Mn d orbitals, we set the on-site Hubbard U
parameter to 2.3 eV~\cite{SIC22}. The van der Waals interaction was modeled using the semiempirical Grimme-D2 correction~\cite{G06,BCF+08}, and a vacuum of 20~\AA~in the $z$-direction to detach the periodic images of the heterostructure was used. The positions of atoms were relaxed using the quasi-Newton scheme using scalar-relativistic pseudopotentials, keeping the force and energy convergence thresholds for ionic minimization to $1\times10^{-4}$~Ry/bohr and $10^{-7}$~Ry, respectively. For the self-consistent calculation, including the spin-orbit coupling, we use the Perdew-Zunger exchange-correlation functional~\cite{PZ81} within the projector augmented-wave method~\cite{pslib,PAW}. Also, we have kept
the same $k$-mesh but increasing the energy convergence thresholds to $10^{-8}$~Ry. Additionally, dipole correction~\cite{B99} was applied to properly determine the Dirac point energy offset due to dipole electric field effects between top/bottom graphene and MnS.

\section{Fitting procedure and effective model parameters of graphene(s)}
In Table 1 of the main text, we report effective parameters for the top and bottom graphene layers in the graphene/MnS/graphene heterostructure at electric field strengths $E=\pm1$~V/nm. These were obtained by fitting the band structure around the $K=4\pi/3a(-1,0)$ and time-reversal symmetric $K'=4\pi/3a(1,0)$ points of graphene (along the $\Gamma K M$ and $\Gamma K' M$ lines, all lying on $k_y=0$). The fitting range extended to a maximal distance of $\pm0.003A^{-1}$ from the K/K' points,
corresponding roughly to a [-30,30]\, meV energy window around the Fermi level. 

\begin{figure}[t]
    \centering
 \includegraphics[width=0.99\linewidth]{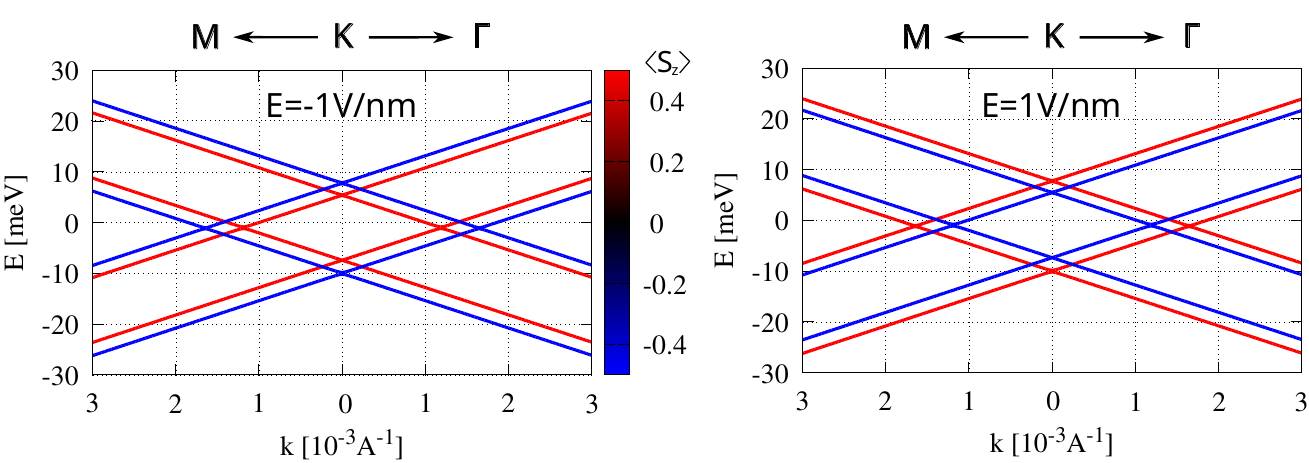}
    \caption{Band structure of graphene/MnS/graphene heterostructure under an applied out-of-plane electric fields of E=$-$1\,V/nm (left panel) and E=1\,V/nm (right panel) in the nonrelativistic case, plotted along the graphene MK$\Gamma$ path. Whereas k describes the distance from the Dirac K point of graphene, the color scale indicates the expectation value of the spin 
    along the (undefined) quantization axis. }
    \label{fig:transport1}
\end{figure}

\subsection{Nonrelativistic approximation}\label{nonrelativistic}

In antiferromagnets with nonrelativistic spin splitting (NRSS AFMs), spin splitting occurs purely from non-relativistic exchange interactions, independently of spin-orbit coupling. To prove that the dominant source of spin-splitting in graphene is not of the relativistic origin, but it comes from the electric-field induced NRSS in MnS, which is transferred to top and bottom graphene by the proximity effect,
we perform a series collinear spin-polarized calculations
for different electric field strengths, $E=\pm0.25$ V/nm, $\pm0.5$ V/nm, and $E=\pm0.75$ V/nm, and $E=\pm1$ V/nm.

In FIG.~S1, we present the band structure of graphene/MnS/graphene heterostructure under an applied out-of-plane electric fields of E=$-$1\,V/nm (left panel) and E=1\,V/nm (right panel) in the nonrelativistic case, plotted along the MK$\Gamma$ path of graphene's Brillouin zone, in the small energy window $[-30,30]$\,meV around the Fermi level. 

The spin-split band structures of the top and bottom graphene layers clearly demonstrate that the observed splitting arises from proximity-induced NRSS originating in the MnS layer. 
A fit of the Hamiltonian model
of top and bottom graphene, $H_{\rm gr/MnS/gr}^{\rm eff}=H_{\rm gr}^{\rm top}+H_{\rm gr}^{\rm bottom}$, to the DFT data
(note that the nonrelativistic approximation requires neglecting the spin-orbit coupling terms in the graphene model defined in the main text) reveals opposite 
ferromagnetic exchange in top versus bottom graphene ($\Delta_{\rm A/B}^{\rm t}>0$, $\Delta_{\rm A/B}^{\rm b}<0$), with the ferromagnetic exchange magnitude varying with the electric field strength across all tested values ($\pm0.25\,$V/nm  to $\pm1$\,V/nm), as one can see in Table~\ref{TAB:NONRELparameters1}. This field-dependent spin splitting, triggered without any spin-orbit coupling rules out the relativistic origins of the splitting and confirms the nonrelativistic proximity transfer to graphene.

In addition, Table~\ref{TAB:magmom} presents the electric-field dependence of averaged magnetic moments on top/bottom C, Mn, and S atoms. The moments on top (Mn$\uparrow$) and bottom (Mn$\downarrow$) Mn atoms differ only marginally across $E \in [-1,1]$~V/nm, confirming that the electric field induces NRSS in MnS without net magnetization or ferrimagnetism in the heterostructure.

\subsection{Relativistic approximation}
In the relativistic case, analyzing the band structure in the vicinity of both the K and K' points of graphene's Brillouin zone is crucial to distinguish spin-orbit coupling parameters (which respect time-reversal symmetry) from magnetic exchange effects (which break it).

A procedure of fitting the DFT data (band energies and spin expectation values along these $k$-paths) to the model described in the main text provides us the effective parameters. In Table~\ref{TAB:RELparameters1} we provide these for field strengths: $E=\pm0.25$ V/nm, $\pm0.5$ V/nm, and $E=\pm0.75$ V/nm. The results confirm the main text conclusions: both top and bottom graphene acquire ferromagnetic proximity exchange ($\Delta_{\rm A/B}^{\rm t}>0$, $\Delta_{\rm A/B}^{\rm b}<0$), with the sign depending on proximity to the ferromagnetic Mn$\uparrow$ ($\Delta_{\rm A/B}^{\rm t}>0$) or Mn$\downarrow$ ($\Delta_{\rm A/B}^{\rm b}<0$) plane. Relative to the meV-scale exchange parameters $\Delta_{\rm A/B}^{\rm t/b}$,
the Rashba SOC terms $\lambda_{\rm R}^{\rm t/b}$ are negligible, while the intrinsic SOC terms $(\lambda_{\rm I}^{\rm A/B})^{\rm t/b}$ act as small perturbations. Additionally, the chemical potential sign in top/bottom graphene correlates directly with the electric field sign (see Table~\ref{TAB:RELparameters1}), as discussed in the main text.

\begin{table}[htp]
\caption{Parameters of the effective tight-binding model of top and bottom graphene in the nonrelativistic case for electric field strengths  E=$\pm$0.25, $\pm$0.5, $\pm$0.75, and $\pm$1\,V/nm.}\label{TAB:NONRELparameters1}
\centering
\small
\setlength{\tabcolsep}{7pt}
\renewcommand{\arraystretch}{1.0}
\begin{tabular}{rrr||rrr}
\hline\hline
E=-1V/nm & top Gr& bottom Gr&E=1V/nm & top Gr& bottom Gr \\ \hline
$v_{\rm F}$ [$10^{6}${\rm m/s}]        &0.819 &0.819 &  $v_{\rm F}$ [$10^{6}${\rm m/s}]       &0.820&0.818\\
$\mu$\;[{\rm meV}]                     &-8.688 &6.604&  $\mu$\;[{\rm meV}]                     & 6.631&-8.663\\
$\Delta$\;[{\rm meV}]                  &-0.085 &-0.081& $\Delta$\;[{\rm meV}]                  &-0.081&-0.082\\
$\Delta_{\rm A}$\;[{\rm meV}]          &1.372 &-1.170&  $\Delta_{\rm A}$\;[{\rm meV}]         &1.115&-1.407\\
$\Delta_{\rm B}$\;[{\rm meV}]          &1.206 &-1.208&  $\Delta_{\rm B}$\;[{\rm meV}]          &1.150& -1.248\\\hline\hline
E=-0.75V/nm & top Gr& bottom Gr&E=0.75V/nm & top Gr& bottom Gr \\ \hline
$v_{\rm F}$ [$10^{6}${\rm m/s}]        &0.818&0.820&   $v_{\rm F}$ [$10^{6}${\rm m/s}]         &0.820&0.818\\
$\mu$\;[{\rm meV}]                     &-7.491&4.707&  $\mu$\;[{\rm meV}]                     &4.954&-7.787\\
$\Delta$\;[{\rm meV}]                  &-0.104&-0.069& $\Delta$\;[{\rm meV}]                 &-0.071&-0.097\\
$\Delta_{\rm A}$\;[{\rm meV}]          &1.300&-1.132&  $\Delta_{\rm A}$\;[{\rm meV}]          &1.097&-1.367\\
$\Delta_{\rm B}$\;[{\rm meV}]          &1.174&-1.200&  $\Delta_{\rm B}$\;[{\rm meV}]          &1.157&-1.232\\\hline\hline
E=-0.5V/nm &  top Gr& bottom Gr&E=0.5V/nm &  top Gr& bottom Gr \\ \hline
$v_{\rm F}$ [$10^{6}${\rm m/s}]        &0.818&0.820&   $v_{\rm F}$ [$10^{6}${\rm m/s}]        &0.820&0.818\\
$\mu$\;[{\rm meV}]                     &-6.142&2.836&  $\mu$\;[{\rm meV}]                     &3.423&-6.516\\
$\Delta$\;[{\rm meV}]                  &-0.143&-0.071& $\Delta$\;[{\rm meV}]                  &-0.056&-0.126\\
$\Delta_{\rm A}$\;[{\rm meV}]          &0.972&-0.869&  $\Delta_{\rm A}$\;[{\rm meV}]          &0.944&-1.171\\
$\Delta_{\rm B}$\;[{\rm meV}]          &0.899&-0.933&  $\Delta_{\rm B}$\;[{\rm meV}]          &1.034 &-1.085\\\hline\hline
E=-0.25V/nm &  top Gr& bottom Gr&E=0.25V/nm &  top Gr& bottom Gr \\ \hline
$v_{\rm F}$ [$10^{6}${\rm m/s}]        &0.818&0.820&    $v_{\rm F}$ [$10^{6}${\rm m/s}]    &0.820&0.818\\
$\mu$\;[{\rm meV}]                     &-5.557&2.070&   $\mu$\;[{\rm meV}]                 &2.537&-5.907\\
$\Delta$\;[{\rm meV}]                  &-0.170&-0.099&  $\Delta$\;[{\rm meV}]              &-0.080&-0.152\\
$\Delta_{\rm A}$\;[{\rm meV}]          &0.457&-0.443&  $\Delta_{\rm A}$\;[{\rm meV}]       & 0.733&-0.878\\
$\Delta_{\rm B}$\;[{\rm meV}]          &0.426&-0.458&   $\Delta_{\rm B}$\;[{\rm meV}]      &0.775&-0.827\\\hline\hline
\end{tabular}
\end{table}

\begin{table}[htp]
\caption{Averaged magnetic moments on top and bottom C, Mn, and S atoms for electric field strengths E=$\pm$1,  E=$\pm$0.75, $\pm$0.5, and $\pm$0.25\,V/nm in the nonrelativistic case.}\label{TAB:magmom}
\centering
\small
\setlength{\tabcolsep}{7pt}
\renewcommand{\arraystretch}{1.0}
\begin{tabular}{rrrrrrr}
\hline\hline
E [V/nm] & C$^{\rm t}$ [$\mu_{\rm B}$]& C$^{\rm b}$ [$\mu_{\rm B}$]&Mn$^{\rm t}$ [$\mu_{\rm B}$]& Mn$^{\rm b}$ [$\mu_{\rm B}$]& S$^{\rm t}$ [$\mu_{\rm B}$]& S$^{\rm b}$ [$\mu_{\rm B}$] \\ \hline
-1 & 1.7$\times10^{-4}$&     -2.1$\times10^{-4}$&4.10848 & -4.10867& 0.03153& -0.03149 \\ \hline
-0.75 &  1.6$\times10^{-4}$& -1.4$\times10^{-4}$&4.10848 &-4.10867& 0.03150 & -0.03149 \\ \hline
-0.5 &  1.3$\times10^{-4}$&  -1.6$\times10^{-4}$&4.10848 & -4.10867& 0.03150 &  -0.03149 \\ \hline
-0.25 & 1.3$\times10^{-4}$&  -1.4$\times10^{-4}$&4.10848 & -4.10867&  0.03150 & -0.03149 \\ \hline
0.25 & 1.4$\times10^{-4}$&   -1.4$\times10^{-4}$&4.10848 & -4.10867& 0.03150 & -0.03149 \\ \hline
0.5 & 1.6$\times10^{-4}$&    -1.4$\times10^{-4}$&4.10848 & -4.10867& 0.03150 & -0.03149 \\ \hline
0.75 & 1.8$\times10^{-4}$&   -1.4$\times10^{-4}$& 4.10848 &-4.10867&  0.03147 & -0.03149 \\ \hline
1 & 2.2$\times10^{-4}$&      -1.6$\times10^{-4}$& 4.10848 & -4.10867& 0.03147 & -0.03150 \\ \hline
\end{tabular}
\end{table} 
\begin{table}[htp]
\caption{Parameters of the effective tight-binding model of top and bottom graphene (relativistic case) within graphene/MnS/graphene heterostrucuture for electric field strengths  E=$\pm$0.75, $\pm$0.5, and $\pm$0.25\,V/nm.}\label{TAB:RELparameters1}
\centering
\small
\setlength{\tabcolsep}{7pt}
\renewcommand{\arraystretch}{1.0}
\begin{tabular}{rrr||rrr}
\hline\hline
E=-0.75V/nm & top Gr& bottom Gr&E=0.75V/nm & top Gr& bottom Gr \\ \hline
$v_{\rm F}$ [$10^{6}${\rm m/s}]        &0.818&0.820&$v_{\rm F}$ [$10^{6}${\rm m/s}]         &0.820&0.818\\
$\mu$\;[{\rm meV}]                     &-7.066&4.536&$\mu$\;[{\rm meV}]                     &4.816&-7.233\\
$\Delta$\;[{\rm meV}]                  &-0.001&-0.054&$\Delta$\;[{\rm meV}]                 &-0.038&0.101\\
$\lambda_{\rm I}^{\rm A}$\;[{\rm meV}] &0.120&-0.016&$\lambda_{\rm I}^{\rm A}$\;[{\rm meV}] &0.054&-0.017\\
$\lambda_{\rm I}^{\rm B}$\;[{\rm meV}] &0.095&0.037&$\lambda_{\rm I}^{\rm B}$\;[{\rm meV}]  &-0.044&0.010\\
$\lambda_{\rm R}$\;[{\rm meV}]         &0.002&0.003&$\lambda_{\rm R}$\;[{\rm meV}]          &-0.002&-0.002\\
$\Delta_{\rm A}$\;[{\rm meV}]          &1.204&-1.038&$\Delta_{\rm A}$\;[{\rm meV}]          &1.011&-1.202\\
$\Delta_{\rm B}$\;[{\rm meV}]          &1.193&-1.209&$\Delta_{\rm B}$\;[{\rm meV}]          &1.148&-1.331\\
\hline\hline
E=-0.5V/nm &  top Gr& bottom Gr&E=0.5V/nm &  top Gr& bottom Gr \\ \hline
$v_{\rm F}$ [$10^{6}${\rm m/s}]        &0.818&0.820&$v_{\rm F}$ [$10^{6}${\rm m/s}]         &0.820&0.818\\
$\mu$\;[{\rm meV}]                     &-6.527&3.211&$\mu$\;[{\rm meV}]                     &3.132&-6.241\\
$\Delta$\;[{\rm meV}]                  &-0.132&-0.059&$\Delta$\;[{\rm meV}]                 &-0.034&0.135\\
$\lambda_{\rm I}^{\rm A}$\;[{\rm meV}] &0.006&-0.015&$\lambda_{\rm I}^{\rm A}$\;[{\rm meV}] &0.012&-0.014\\
$\lambda_{\rm I}^{\rm B}$\;[{\rm meV}] &-0.014&0.016&$\lambda_{\rm I}^{\rm B}$\;[{\rm meV}] &-0.008&0.006\\
$\lambda_{\rm R}$\;[{\rm meV}]         &0.001&0.001&$\lambda_{\rm R}$\;[{\rm meV}]          &-0.001&-0.001\\
$\Delta_{\rm A}$\;[{\rm meV}]          &1.102&-0.976&$\Delta_{\rm A}$\;[{\rm meV}]          &0.842&-1.015\\
$\Delta_{\rm B}$\;[{\rm meV}]          &1.001&-1.057&$\Delta_{\rm B}$\;[{\rm meV}]     &0.985&-1.092\\\hline\hline
E=-0.25V/nm &  top Gr& bottom Gr&E=0.25V/nm &  top Gr& bottom Gr \\ \hline
$v_{\rm F}$ [$10^{6}${\rm m/s}]        &0.818&0.826&$v_{\rm F}$ [$10^{6}${\rm m/s}]          &0.820&0.818\\
$\mu$\;[{\rm meV}]                     &-5.513&2.437&$\mu$\;[{\rm meV}]                      &2.603&-5.489\\
$\Delta$\;[{\rm meV}]                  &-0.168&-0.017&$\Delta$\;[{\rm meV}]                  &-0.089&0.160\\
$\lambda_{\rm I}^{\rm A}$\;[{\rm meV}] &0.001&-0.011&$\lambda_{\rm I}^{\rm A}$\;[{\rm meV}]  &0.001&-0.009\\
$\lambda_{\rm I}^{\rm B}$\;[{\rm meV}] &-0.008&-0.002&$\lambda_{\rm I}^{\rm B}$\;[{\rm meV}] &-0.010&0.003\\
$\lambda_{\rm R}$\;[{\rm meV}]         &0.001&0.000&$\lambda_{\rm R}$\;[{\rm meV}]           &0.000&-0.001\\
$\Delta_{\rm A}$\;[{\rm meV}]          &0.628& -0.587&$\Delta_{\rm A}$\;[{\rm meV}]          &0.620&-0.713\\
$\Delta_{\rm B}$\;[{\rm meV}]          &0.620&-0.678&$\Delta_{\rm B}$\;[{\rm meV}]      &0.660&-0.766\\\hline\hline
\end{tabular}
\end{table}

\begin{figure}[t]
    \centering
\includegraphics[width=0.7\linewidth]{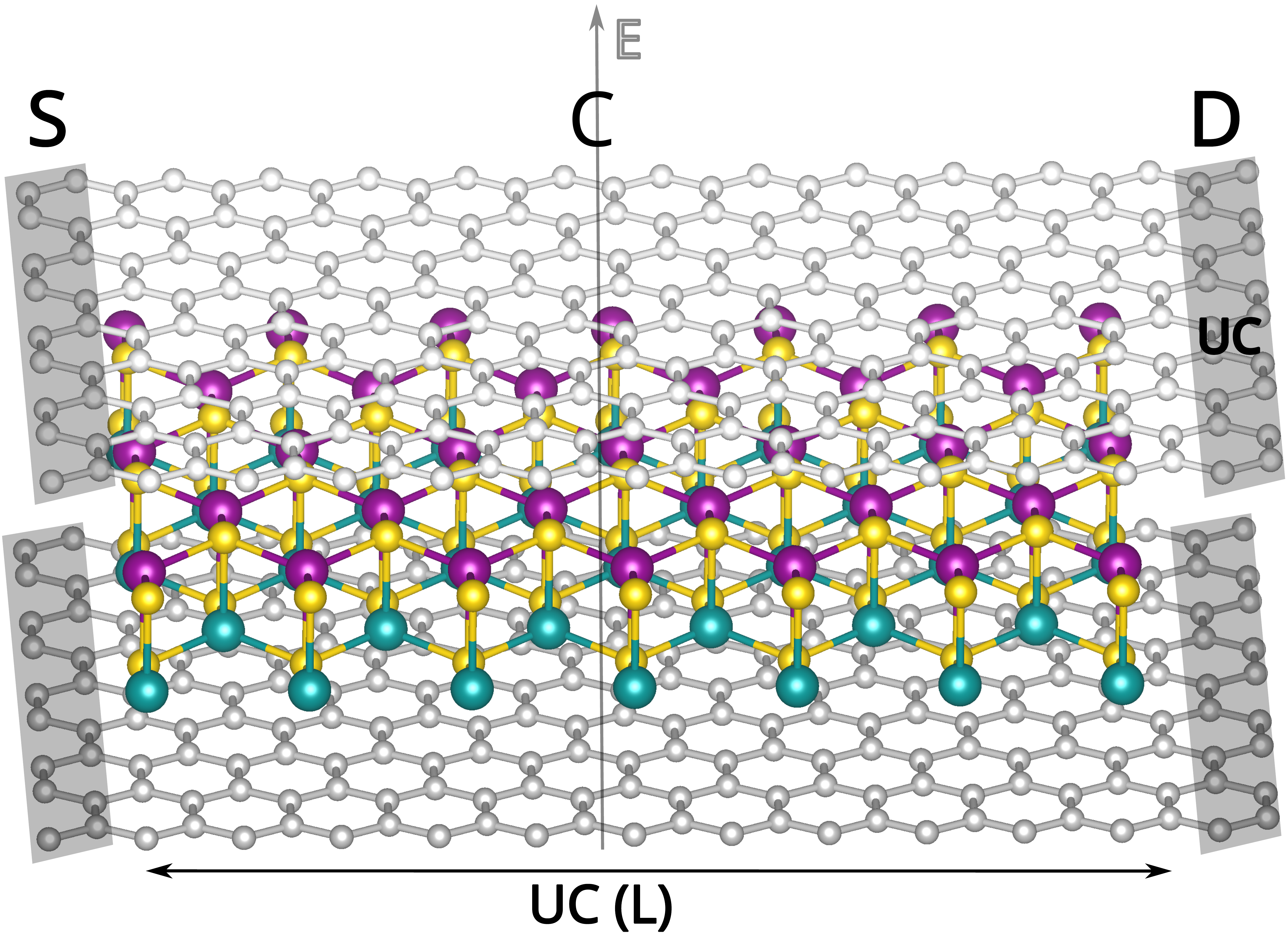}
    \caption{
  Detailed view of the transport device which implements the 
   proximity-induced exchange in graphene/MnS/graphene heterostructure. The source and drain are labeled with S and D, respectively, consisting on free-standing top and bottom graphene. The central region $C$ consists of graphene/MnS/graphene under applied perpendicular electric field $E$, in which top and bottom graphene are proximitized by an NRSS AFM. Whereas the setup is illustrated on the central part consisting of $L=11$ repetitions of the elementary units cell (UC) that has 18 carbon atoms, we have studied the impact of the proximity effect on conductance of nanoribbons with different length, labeled with L. In all cases, the width of the nanoribbon is kept at 17.8\,\AA.  
   }
    \label{fig:transportAPP}
\end{figure}

\section{Transport properties of a graphene/MnS/graphene heterostructure with (graphene) zigzag edges}

To probe the transport signatures of NRSS in the studied graphene/MnS/graphene heterostructure, we study zigzag-edged graphene nanoribbons where pristine source and drain regions are interfaced with a central synthetic NRSS AFM. Whereas in FIG. 1 of the main text we have presented a schematic view of the transport device, in FIG.~\ref{fig:transportAPP} we present the detailed view of the given device, in which the length of the nanoribbon is given in terms of number $L$ of the unitary cells (UCs). We note that the length of the nanoribbon studied in  the main text corresponds to $L=20$.    

\begin{figure}[t]
    \centering
 \includegraphics[width=0.99\linewidth]{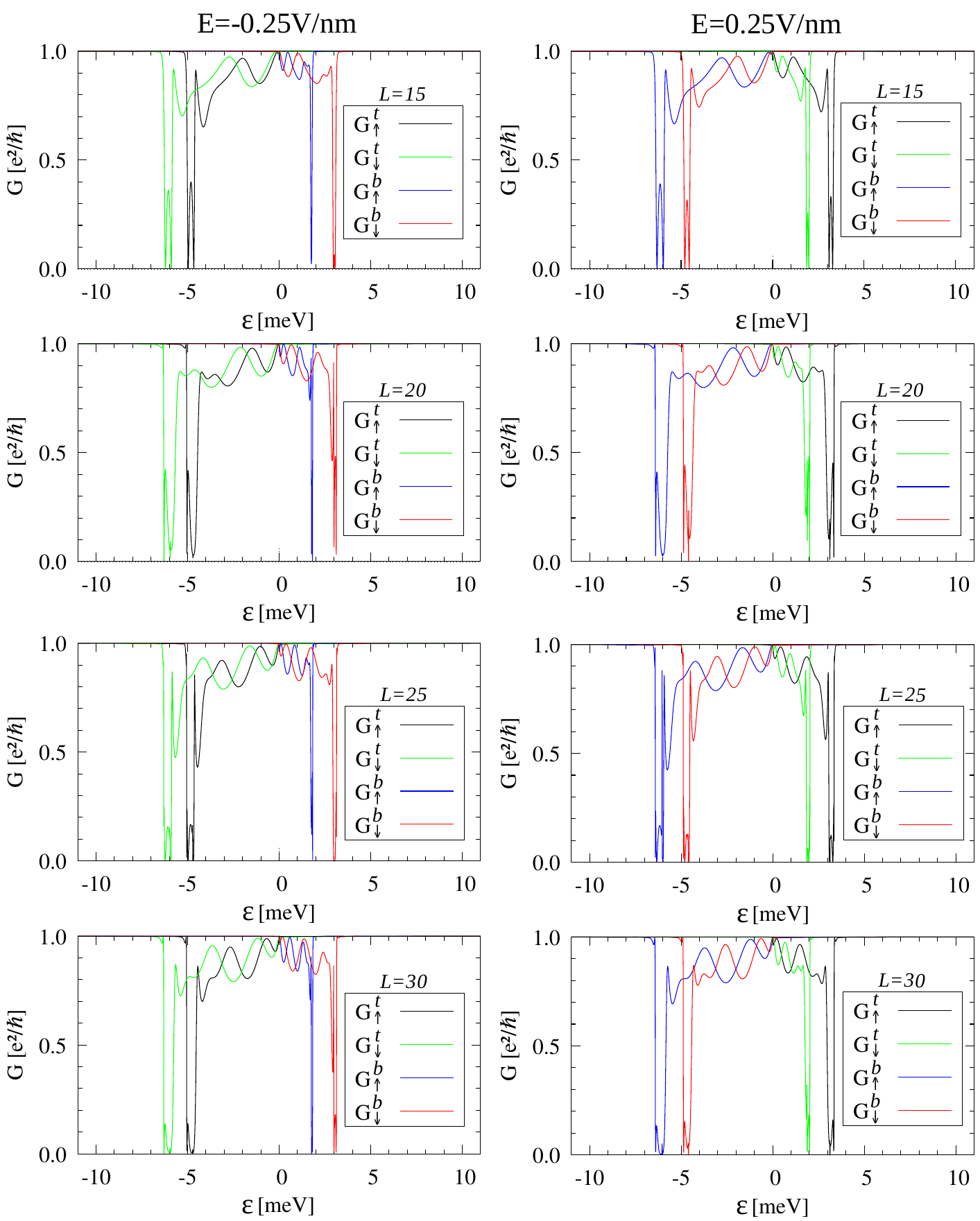}
    \caption{Conductance $G_\sigma^{t/b}(\varepsilon)$ vs.~energy $\varepsilon$ for graphene nanoribbons of different lengths $L=15,20,25$, and $30$. Electric field strengths studied: $E=-0.25/0.25$ V/nm (left/right panels).}
    \label{fig:transport025}
\end{figure}

\begin{figure}[t]
    \centering
 \includegraphics[width=0.99\linewidth]{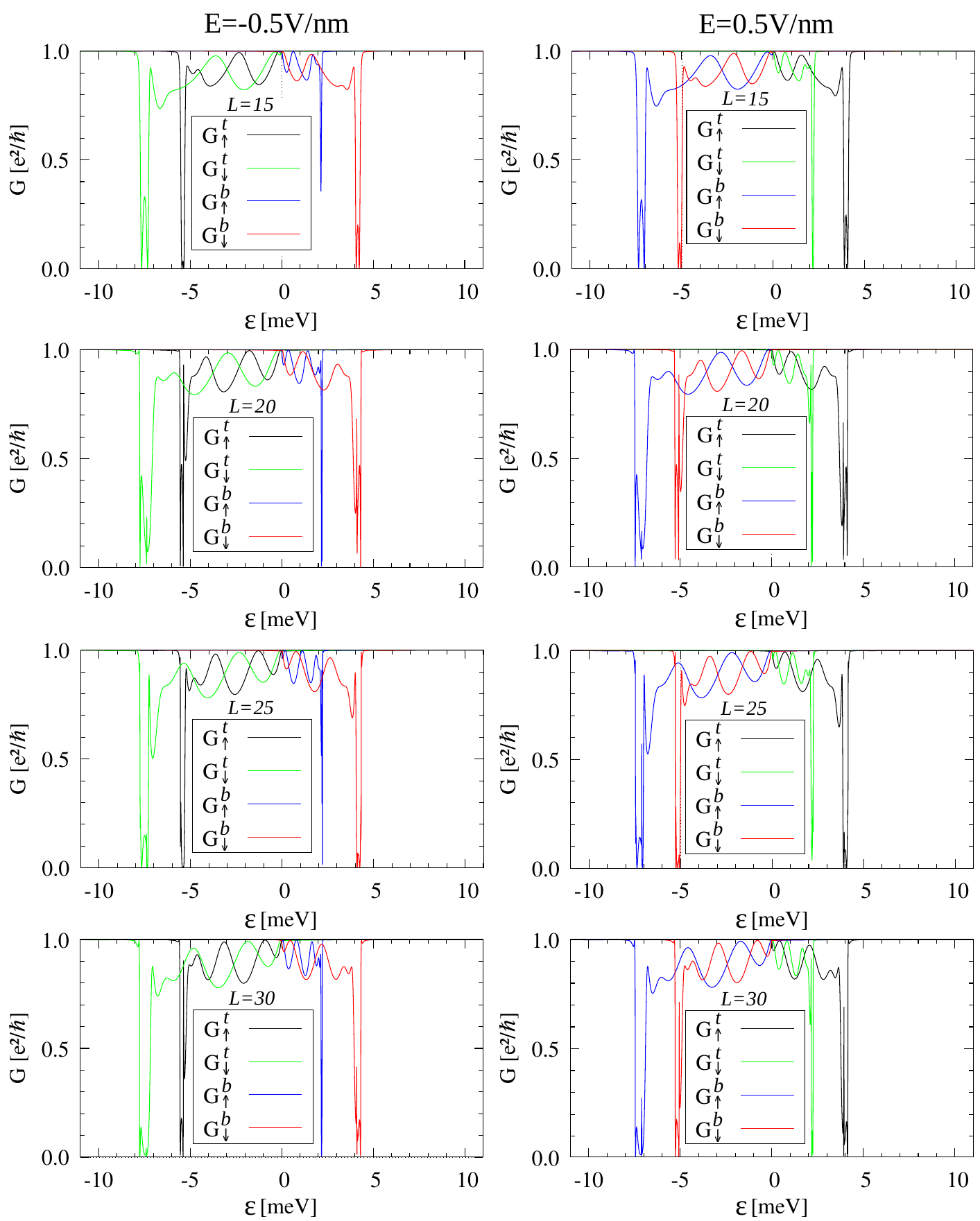}
    \caption{Conductance $G_\sigma^{t/b}(\varepsilon)$ vs.~energy $\varepsilon$ for graphene nanoribbons of different lengths $L=15,20,25$, and $30$. Electric field strengths studied: $E=-0.5/0.5$ V/nm (left/right panels).}    \label{fig:transport05}
\end{figure}

\begin{figure}[htp]
    \centering
 \includegraphics[width=0.99\linewidth]{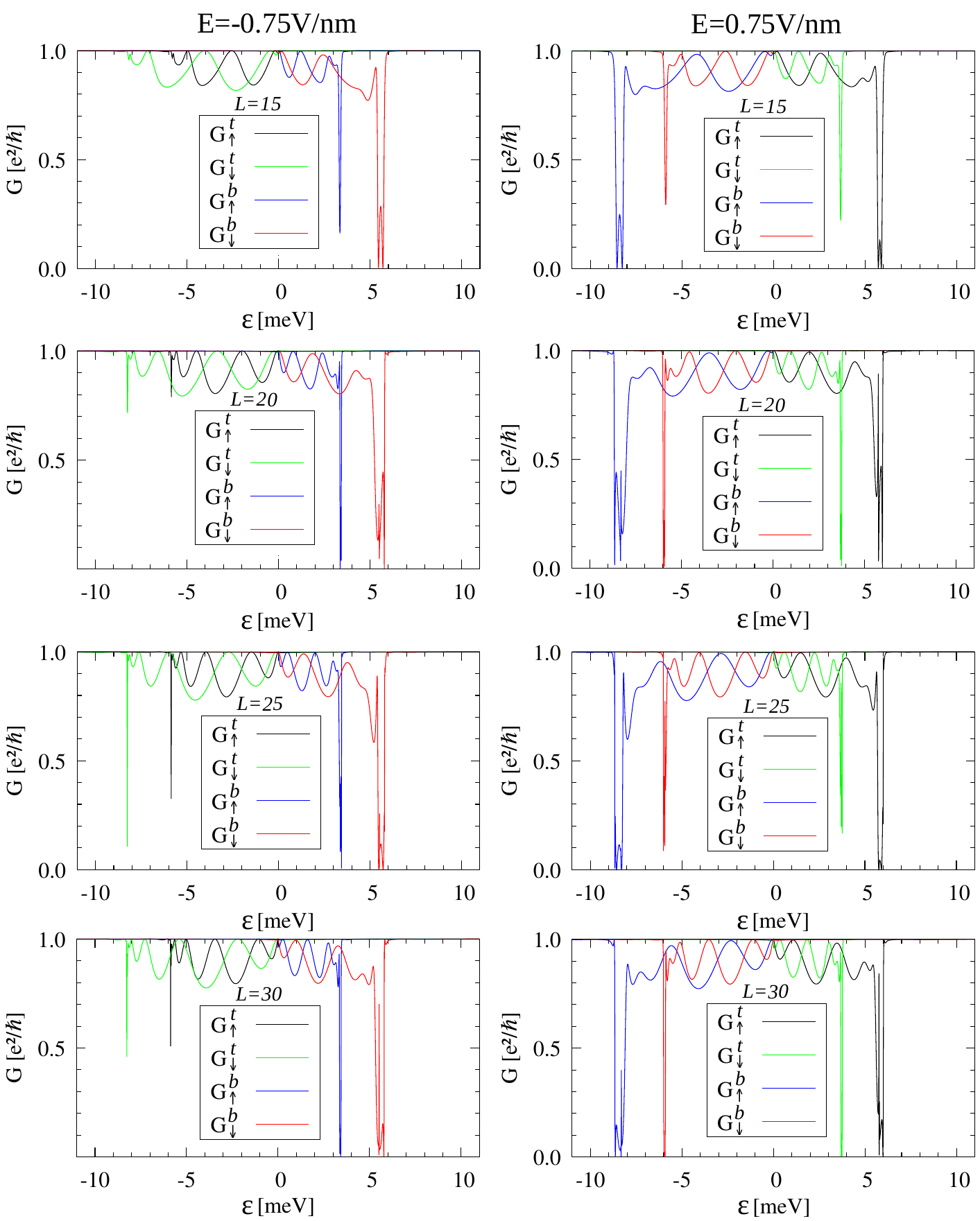}
    \caption{Conductance $G_\sigma^{t/b}(\varepsilon)$ vs.~energy $\varepsilon$ for graphene nanoribbons of different lengths $L=15,20,25$, and $30$. Electric field strengths studied: $E=-0.75/0.75$ V/nm (left/right panels).}
    \label{fig:transport075APP}
\end{figure}

\begin{figure}[htp]
    \centering
 \includegraphics[width=0.99\linewidth]{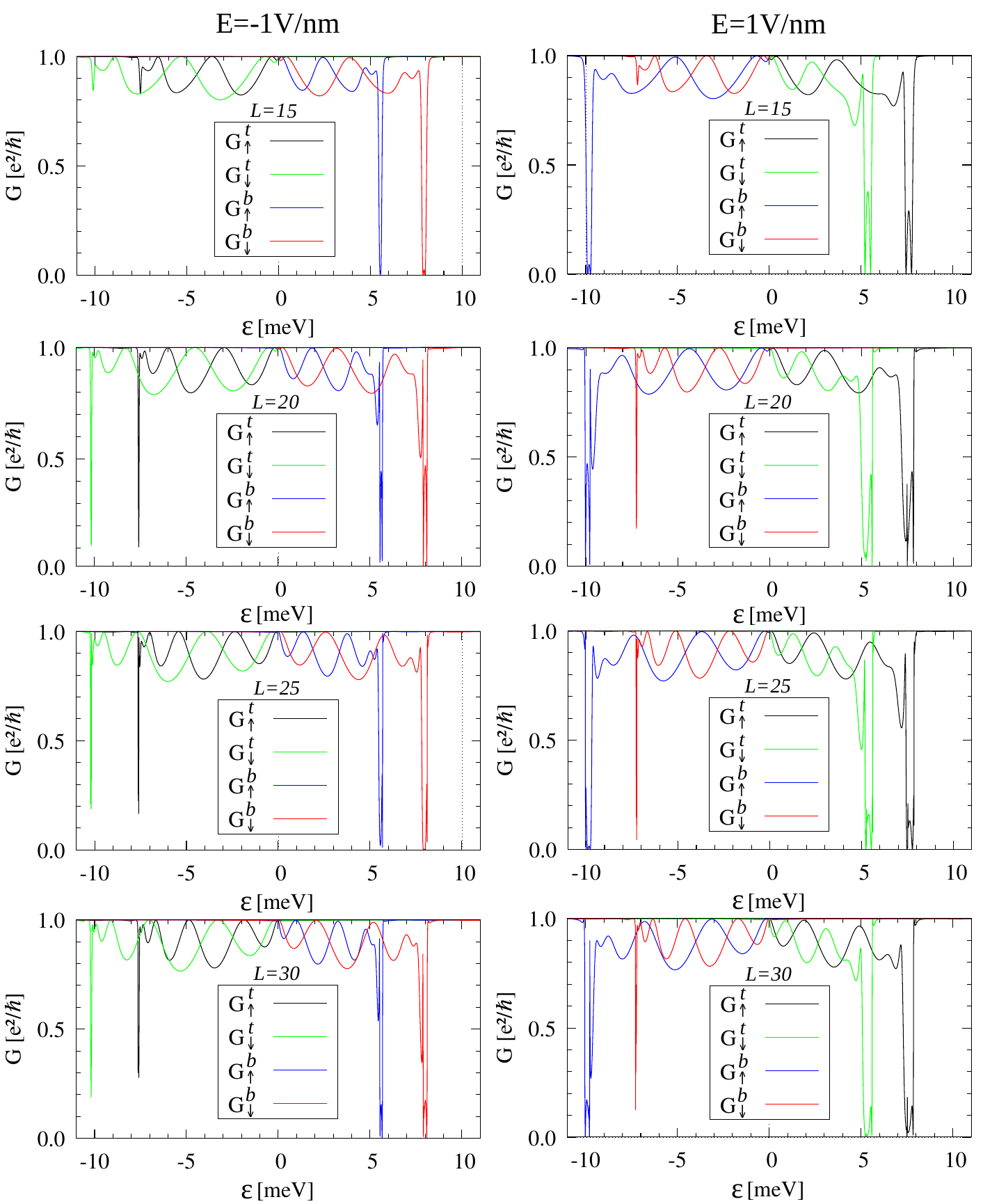}
    \caption{Conductance $G_\sigma^{t/b}(\varepsilon)$ vs.~energy $\varepsilon$ for graphene nanoribbons of different lengths $L=15,20,25$, and $30$. Electric field strengths studied: $E=-1/1$ V/nm (left/right panels).}\label{fig:transport1APP}
\end{figure}

In FIGS.~\ref{fig:transport025}-\ref{fig:transport1APP} we present the spin-resolved conductances $G^{\rm t/b}_{\uparrow\downarrow}$ as a function of energy $\varepsilon$ for graphene nanoribbons with different lengths $L = 15, 20, 25,$ and $30$ at electric field strengths $E = \pm 0.25$ V/nm (FIG.~\ref{fig:transport025}); $E = \pm 0.5$ V/nm (FIG.~\ref{fig:transport05}); $E = \pm 0.75$ V/nm (FIG.~\ref{fig:transport075APP}); and $E = \pm 1$ V/nm (FIG.~\ref{fig:transport1APP}). Calculations confirm conductance dips in the spin-resolved top and bottom graphene channels. These dips appear within a narrow energy window $|\varepsilon| \leq 11$ meV around the Fermi level of the leads, arising from sublattice-resolved ferromagnetic proximity exchange with opposite signs in top/bottom layers and electric-field-tuned doping. These features sharpen and deepen with increasing central region length $L$ due to finite-size effects, whereas the presence of two visible dips with small energy difference steam from sublattice exchange differences ($\Delta_{\rm A}^{\rm t/b}\neq \Delta_{\rm B}^{\rm t/b}$), while staggered potentials $\Delta^{\rm t/b}$ introduce asymmetries between spin channels. The effect strengthens with electric field strength $|E|$, as larger fields enhance exchange parameters and chemical potential shifts $\mu^{\rm t/b}$, driving larger deviations from ideal conductance, being equal to $e^2/\hbar$ in the analyzed energy range~\cite{YZS+09}. 

We note that the $\mathcal{RT}$ symmetry relating the conductance profiles at $+E$ and $-E$ is only approximate in the present heterostructure, since the asymmetric electrostatic screening of the applied field by the MnS layer induces slightly different charge redistributions in
the top and bottom graphene layers depending on the field
direction, as reflected in the tight-binding parameters of
Table~\ref{TAB:RELparameters1} and the averaged magnetic moments of Table~\ref{TAB:magmom},
which are not exactly interchanged upon field reversal.


\bibliographystyle{unsrt}
\bibliography{biblio}